\documentclass[12pt]{iopart}

\usepackage[dvips]{graphicx}
\usepackage{bm}

\begin{document}

\sloppy

% \title{Coulomb Interactions via Local Dynamics: \\
% A Molecular--Dynamics Algorithm}
\title[Coulomb Interactions via local MD]
{Coulomb Interactions via Local Dynamics: \\
A Molecular--Dynamics Algorithm}

% \author{Igor Pasichnyk}
% \author{Burkhard D\"unweg}
% \affiliation{Max Planck Institute for Polymer Research, \\
% Ackermannweg 10, D--55128 Mainz, Germany}

\author{Igor Pasichnyk and Burkhard D\"unweg}
\address{Max Planck Institute for Polymer Research, \\
Ackermannweg 10, D--55128 Mainz, Germany}

% \date{\today}

\begin{abstract}
  We derive and describe in detail a recently proposed method for
  obtaining Coulomb interactions as the potential of mean force
  between charges which are dynamically coupled to a local
  electromagnetic field.  We focus on the Molecular Dynamics version
  of the method and show that it is intimately related to the
  Car--Parrinello approach, while being equivalent to solving
  Maxwell's equations with freely adjustable speed of light.
  Unphysical self--energies arise as a result of the lattice
  interpolation of charges, and are corrected by a subtraction scheme
  based on the exact lattice Green's function. The method can be
  straightforwardly parallelized using standard domain decomposition.
  Some preliminary benchmark results are presented.
\end{abstract}


\maketitle

\section{Introduction}
\label{sec:intro}

Simulations of charged systems face a big computational challenge due
to the long--range nature of the electrostatic interaction. If $N$ is
the number of charges, then the computational cost of the most naive
approach to evaluate the interaction energy would scale as $N^2$,
since every charge interacts with every other charge. Very
sophisticated algorithms have been developed to tackle this problem
and to reduce the computational complexity. The most prominent ones
are the so--called P$^3$M method (``particle--particle /
particle--mesh''), which is based on Fast Fourier Transforms and
scales as $N \log N$ \cite{hockney}, and the Fast Multipole method
\cite{greengard} which scales linearly with $N$.

A similar problem arises in the simulation of Brownian particles which
interact hydrodynamically: Their stochastic displacements are highly
correlated, due to fast diffusive momentum transport through the
solvent. For sufficiently slow particles, a quasi--static
approximation works excellently, and in this case the correlation
function decays as $1 / r$ ($r$ interparticle distance)
\cite{doiedw}, just as in electrostatics. For these systems,
it has turned out that it is both much simpler and also more
efficient to explicitly simulate the momentum transfer through
the surrounding solvent. This makes the simulation of
several ten thousands of Brownian particles feasible
\cite{laddprl,patrickscreening}. Although most of the
computational effort goes into the flow field (for {\em two} reasons
--- one needs reasonable spatial resolution of the flow field, and it
moves much faster than the Brownian particles), this approach
ultimately wins, because it is inherently local, and therefore scales
linearly with $N$.

This observation raises the question if something similar could be
tried for Coulomb interactions. After all, electrostatics is just the
quasi--static limit of full electrodynamics. The obvious approach
would be to couple a system of charges to an electromagnetic field
which propagates according to the Maxwell equations (ME), and then run
Molecular Dynamics (MD). A suitable acronym for such a method might be
MEMD (``Maxwell equations Molecular Dynamics''). Just as in the
hydrodynamic case, this is an intrinsically local algorithm, and
therefore scales linearly. The instantaneous $1/r$ interaction is thus
replaced by some retarded interaction travelling with the speed of
light $c$. Using the actual physical value of $c$ will of course not
work, since then the separation of time scales between charges and
fields will be prohibitive. However, there is no need to take such a
large $c$ value. It is sufficient to just make $c$ large enough such
that the quasi--static approximation still holds to sufficient
accuracy. This is the lesson we have learned from Car--Parrinello (CP)
simulations \cite{carparrinello}, where the electrons are assigned an
unphysically large mass, precisely for the same reason. The analogy
between MEMD and CP actually goes much further, as we will see
below. This should not be too much of a surprise, since the universal
applicability of the CP approach to a wide variety of problems in
physics (e.~g. classical field theories) has already been observed in
the original publication \cite{carparrinello}, and exploited in the
context of classical density--functional theory \cite{loewen}.

The MEMD idea has been pursued recently by A. C. Maggs and
collaborators \cite{maggs_prl_1,maggs_jcp,maggs_prl_2}, and by us, in
close contact with him. He has made a couple of very important
observations, which have deepened our insight into the approach
significantly, and contributed to the answer of a number of very
important questions:
\begin{enumerate}
\item Is Maxwell dynamics the only possible
      way to propagate the fields? The answer is no; it is
      also possible to propagate them in a {\em diffusive}
      fashion. This has been implemented by means of a
      Monte Carlo algorithm \cite{maggs_prl_1,maggs_jcp}
      for a lattice gas of charges.
\item If we restrict attention to Hamiltonian or quasi--Hamiltonian
      dynamics of the system, and want wave--like propagation of the
      signal, is then Maxwell--style dynamics
      the only choice? The answer is a cautious yes; one can
      show that the Maxwell equations arise in a very natural
      way if one derives the method along the lines of CP.
\item Is there a contradiction between the Lorentz covariance
      of the ME, and the strictly nonrelativistic setup of MD?
      The answer is no; the Lorentz covariance actually has to
      do with the fact that the value of $c$ is the same in all
      reference frames. This, however, is not the case here:
      In our context, $c$ means nothing but the propagation
      velocity of electromagnetic waves {\em relative to the
      discretization lattice} which provides an absolute
      reference frame (an ``ether'').
\item Is it necessary to use a large value of $c$ to avoid
      violation of a quasi--static behavior? The answer
      is no as long as just static properties of the system
      in thermal equilibrium are considered --- the values
      of these properties turn out to be completely independent
      of $c$.
\item Is it necessary to apply a thermostat to the system?
      Ref. \cite{maggs_prl_2} claims yes, in order to avoid
      unwanted conserved quantities. Our belief is no, based
      upon the fact that the particle dynamics provides lots
      of nonlinearities into the equations of motion. For
      more details, see below.
\item How is MEMD implemented? The previous papers have been
      rather brief on this issue; we try to provide somewhat
      more detail.
\item How does MEMD perform, in particular in comparison
      with existing methods? In this respect, there is also
      so far only little information available. In the present
      paper, we report some benchmark results which give us
      some feeling for the quality of the approach --- although
      these are quite preliminary, and still far from
      providing a clear and comprehensive picture.
\end{enumerate}
In what follows, we will essentially re--derive the MEMD algorithm put
forward in Ref. \cite{maggs_prl_2}, and discuss some details of our
implementation, which differs slightly from that of Ref.
\cite{maggs_prl_2}. We will then present some benchmark results,
comparing MEMD with P$^3$M for the same system. For our chosen set of
parameters, we find rather similar or even better computational
efficiency. However, this comparison should not yet be considered as
the final answer: Firstly, MEMD can probably still be speeded up
significantly by combining it with a direct evaluation of Yukawa--like
forces on short length scales, roughly along the lines as suggested in
Ref. \cite{maggs_prl_2}. Secondly, the dependence on the thermodynamic
state point (in particular, on density) has not yet been investigated.
Physically, it is clear that the efficiency of MEMD depends (i) on the
number of operations required to propagate the system for one time
step, and (ii) on the time needed to build up electrostatic
correlations on the relevant length scale, which is, in essence, the
Debye screening length. For this reason, one should expect that the
efficiency depends rather strongly on the speed of light, and also on
the density, since the Debye length decreases as a function of
density. In other words, one expects MEMD to work particularly well
for rather dense systems simulated with a large $c$ value --- and this
is precisely the regime where our preliminary comparison has been
done.

\section{Continuum Theory}
\label{sec:theory}

We start out from Maxwell's equations in vacuum, using standard SI
units:
\begin{eqnarray}
\label{eq:maxwell1}
\vec \nabla \cdot \vec E & = & \frac{1}{\epsilon_0} \rho  \\
\label{eq:maxwell2}
\vec \nabla \times \vec E & = & - \frac{1}{\epsilon_0 c^2}
\frac{\partial \vec H}{\partial t} \\
\label{eq:maxwell3}
\vec \nabla \cdot \vec H & = & 0 \\
\label{eq:maxwell4}
\vec \nabla \times \vec H & = & \vec j + \epsilon_0
\frac{\partial \vec E}{\partial t} ,
\end{eqnarray}
where $\epsilon_0$ is the vacuum dielectric constant,
$c$ the speed of light, $\vec E$ the electric field, $\vec H$ the
magnetic field, $\rho$ the charge density and $\vec j$ the current
density, which are coupled via the continuity equation
\begin{equation} \label{eq:continuity}
\frac{\partial \rho}{\partial t} + \vec \nabla \cdot \vec j = 0 .
\end{equation}
Electrostatics is obtained by setting the current and
all time derivatives to zero, implying that the magnetic
field vanishes:
\begin{eqnarray}
\label{eq:statics1}
\vec \nabla \cdot \vec E & = & \frac{1}{\epsilon_0} \rho  \\
\label{eq:statics2}
\vec \nabla \times \vec E & = & 0 .
\end{eqnarray}
Note that this set of equations also results from taking the limit $c
\to \infty$. This means that the electric field will be just an
electrostatic field as long as the charges move at much slower
velocity than $c$. Furthermore, the Lorentz force on a charge $e$,
\begin{equation} \label{eq:lorentzforce}
\vec F_L = e \left( \vec E + \frac{1}{\epsilon_0 c^2} 
\vec v \times \vec H \right)
\end{equation}
($\vec v$ denoting the charge's velocity) will just reduce to the
electrostatic force $e \vec E$ in the same limit. This is the
justification of the fact that MD simulations of common materials
usually treat interactions of charges just as electrostatics. In turn
this means that one will obtain electrostatic behavior whenever $c$ is
large compared to all particle velocities, as already stated in Sec
\ref{sec:intro}.

The conventional approach tries to find a solution to Eqs.
\ref{eq:statics1} and \ref{eq:statics2}, for a given charge
distribution $\rho$. Note that {\em both} equations must be satisfied
for strict electrostatics, since Eq. \ref{eq:statics1} only fixes the
longitudinal component of the electric field, while the condition of
vanishing transversal component is coded in Eq. \ref{eq:statics2}. No
{\em local} way of finding the solution directly is known.

As a first step, we re--formulate the electrostatic problem in terms
of a constrained variational problem. Gauss' law (Eq.
\ref{eq:statics1}) is viewed as a {\em constraint} which selects a
certain surface out of the space of electric field configurations; we
will call this the ``constraint surface'' (CS). We now minimize the
electric field energy,
\begin{equation} \label{eq:electricfieldenergy}
{\cal H}_{EF} = \frac{\epsilon_0}{2} \int d^3 \vec r \, \vec E^2
\end{equation}
under the constraint Eq. \ref{eq:statics1}. This can be done as
follows: Suppose $\vec E_0$ is {\em some} field on the CS, where a
non--vanishing transversal component is admitted.  Then all fields on
the CS can be written in the form
\begin{equation}
\vec E = \vec E_0 + \vec \nabla \times \vec \Theta ,
\end{equation}
where $\vec \Theta$ is allowed to pass through all field configurations
without any restriction. We thus write ${\cal H}_{EF}$ in terms
of the $\vec \Theta$ field,
\begin{equation} \label{eq:electrostatichamiltonian1}
{\cal H}_{EF} \left( \left\{ \vec \Theta \right\} \right) =
\frac{\epsilon_0}{2} \int d^3 \vec r 
\left( \vec E_0 + \vec \nabla \times \vec \Theta \right)^2
\end{equation}
and the minimum condition as
\begin{equation}
\frac{\delta}{\delta \vec \Theta}
{\cal H}_{EF} \left( \left\{ \vec \Theta \right\} \right) = 0
\end{equation}
or
\begin{equation} \label{eq:solutionelectrostaticminimum}
0 = \vec \nabla \times 
\left( \vec E_0 + \vec \nabla \times \vec \Theta \right)
= \vec \nabla \times \vec E,
\end{equation}
i.~e. Eq. \ref{eq:statics2}. The variational problem is thus seen to
be equivalent to the original electrostatic problem. We can say that
the system is on its Born--Oppenheimer surface (BOS) if, for given
charge distribution $\rho$, the electric fields are on the constraint
surface, and the field energy is minimal.

Alternatively, one may also look at the problem in Fourier space: Let
the Fourier transform of $\vec E (\vec r)$ be defined as
\begin{equation}
\tilde{\vec E} (\vec k) = (2 \pi)^{-3/2} \int d^3 \vec r \,
\vec E(\vec r) \exp \left( - i \vec k \cdot \vec r \right) 
\end{equation}
and let $\hat k$ denote the unit vector in the direction of $\vec
k$. Then we can decompose the electric field into a longitudinal
component $\tilde{\vec E}_{\parallel}$ and a transversal component
$\tilde{ \vec E}_{\perp}$, $\tilde{\vec E} = \tilde{\vec
E}_{\parallel} + \tilde{\vec E}_{\perp}$ with $\tilde{\vec
E}_{\parallel} \cdot \hat k = \tilde E_{\parallel}$ and $\tilde{\vec
E}_{\perp} \cdot \hat k = 0$. Then Eqs. \ref{eq:statics1} and
\ref{eq:statics2} are transformed to $i \vec k \cdot \tilde{\vec E}
= \tilde \rho / \epsilon_0$ and $i \vec k \times \tilde{\vec E} = 0$,
or $\tilde E_{\parallel} = \tilde \rho / (i k \epsilon_0)$ and $\tilde
E_{\perp} = 0$. Furthermore, the electric field energy can be written
as
\begin{eqnarray} \label{eq:newelectricfieldenergy}
{\cal H}_{EF} & = &
\frac{\epsilon_0}{2} \int d^3 \vec k \, 
\left\vert \tilde{\vec E} \right\vert^2
=  
\frac{\epsilon_0}{2} \int d^3 \vec k \, 
\left\vert \tilde E_{\parallel} \right\vert^2
+
\frac{\epsilon_0}{2} \int d^3 \vec k \, 
\left\vert \tilde E_{\perp} \right\vert^2
\nonumber
\\
& = &
\frac{1}{2 \epsilon_0} \int d^3 \vec k \,
\frac{\left\vert \tilde \rho \right\vert^2}{k^2}
+
\frac{\epsilon_0}{2} \int d^3 \vec k \, 
\left\vert \tilde E_{\perp} \right\vert^2 .
\end{eqnarray}
Again, one sees that the longitudinal component is determined by the
charge distribution, while the transversal component is just minimized
away.

In the spirit of CP simulations, we now wish to replace the precise
solution of the minimization problem by some (to a certain degree
artificial) dynamics which keeps the system precisely on the CS but
allows fluctuations around the BOS. An interesting observation by
Maggs \cite{maggs_prl_1} is that {\em arbitrarily large} deviations
from the BOS are permitted as long as one does statistical mechanics
in the canonical ensemble, and is interested in static properties
only. This is easily understood by looking at
Eq. \ref{eq:newelectricfieldenergy}: There one sees that the total
Hamiltonian decomposes into two {\em additive} contributions: The
first term $\cal{H}_{\parallel}$ is just the standard electrostatic
Coulomb Hamiltonian, while the second term $\cal{H}_{\perp}$ is the
energy stored in the additional transversal degree of freedom,
describing the amount of deviation from the BOS. Additivity, however,
implies that the Boltzmann factor factorizes,
\begin{equation}
\exp \left( - \beta \cal{H} \right) =
\exp \left( - \beta \cal{H}_{\parallel} \right)
\exp \left( - \beta \cal{H}_{\perp} \right) 
\end{equation}
($\beta = 1 / (k_B T)$, where $T$ is the absolute temperature and
$k_B$ Boltzmann's constant), which in turn means that the (artificial)
transversal degree of freedom is statistically independent from the
physical longitudinal one, and hence does not affect statistical
averages of observables which only depend on the charge configuration.
This will be worked out in some more detail at the end of this
section.

Having relaxed the condition Eq. \ref{eq:statics2}, we now turn our
attention to Eq. \ref{eq:statics1}. Suppose that at time $t= 0$ we
have found the full solution to Eqs. \ref{eq:statics1} and
\ref{eq:statics2} by some (slow) procedure; we call this solution
$\vec E_0 (t = 0)$. This is obviously on the CS. The system will then
stay on the CS if the {\em time derivative} of Gauss' law vanishes:
\begin{equation} \label{eq:timederivativeofgauss}
\vec \nabla \cdot \dot{\vec E} - \frac{1}{\epsilon_0} \dot{\rho} = 0.
\end{equation}
Now as the dynamics proceeds, the continuity equation, Eq.
\ref{eq:continuity}, will automatically hold as long as charges are
moved around in the simulation cell by a local updating scheme. This
allows us to re--write Eq. \ref{eq:timederivativeofgauss} as
\begin{equation} \label{eq:timederivativeofgauss2}
\vec \nabla \cdot \left( \dot{\vec E}
+ \frac{1}{\epsilon_0} \vec j \right) = 0.
\end{equation}
We can therefore use the current density to straightforwardly
construct an electric field which stays on the CS. One just has to
integrate $\dot{\vec E} = - \vec j / \epsilon_0$ in time; this is a
manifestly local updating scheme. We thus obtain
\begin{equation}
\vec E_0 (t) = \vec E_0 (t = 0) - \frac{1}{\epsilon_0}
\int_0^t d \tau \, \vec j (\tau) .
\end{equation}
This solution obviously is on the CS, but unfortunately not the
correct electric field (example: For a constant ring current with
vanishing charge density one would obtain an electric field which
grows linearly in time). We therefore generalize this, as before, to
\begin{equation}
\vec E (t) = \vec E_0 (t) + \vec \nabla \times \vec \Theta (t)
\end{equation}
with $\vec \Theta (t = 0) = 0$. However, we now do not minimize ${\cal
  H}_{EF}$ with respect to $\vec \Theta$, but rather supply some
artificial dynamics to this field. There is no unique way of doing
this. One possibility is to postulate an overdamped relaxational
dynamics governed by ${\cal H}_{EF}$; this has been explored in detail
in Ref. \cite{maggs_jcp}. In the present paper, we rather study, as in
Ref. \cite{maggs_prl_2}, a CP--style dynamics, where the equation of
motion for $\vec \Theta$ is of second order in time. We thus need to
supply an initial condition for $\dot{\vec \Theta}$, too; we choose
$\dot{\vec \Theta} (t = 0) = 0$. The most straightforward way to
generate a coupled dynamics is to add a kinetic energy term $ (1/2)
(\epsilon_0 / c^2) \int d^3 \vec r \, \dot{\vec \Theta}^2$ to the
system Lagrangian; here the prefactor is a mass--like parameter, to be
freely chosen in analogy to the electron mass in CP. $c$ will later on
turn out to be the speed of light.

Since $\vec E_0$ depends on the charge distribution in a not very
straightforward way, it is more convenient to rather write the
Lagrangian in terms of the total field $\vec E$, and to take into
account the integration of $\dot{\vec{E_0}} = - \vec j / \epsilon_0$
by means of a non--holonomic constraint which keeps the system on the
CS:
\begin{equation}
\epsilon_0 \dot{\vec E} = 
\epsilon_0 \dot{\vec{E_0}}
+ \epsilon_0 \vec \nabla \times \dot{\vec \Theta} =
- \vec j
+ \epsilon_0 \vec \nabla \times \dot{\vec \Theta} ,
\end{equation}
i.~e.
\begin{equation} \label{eq:newmaxwell4}
\epsilon_0 \dot{\vec E} + \vec j - 
\epsilon_0 \vec \nabla \times \dot{\vec \Theta} = 0 .
\end{equation}
This is nothing but the fourth Maxwell equation, Eq.
\ref{eq:maxwell4}, if we identify $\vec H = \epsilon_0 \dot{\vec
  \Theta}$. We thus see that the continuity equation (Eq.
\ref{eq:continuity}) as well as the first and fourth Maxwell equation
(Eqs. \ref{eq:maxwell1} and \ref{eq:maxwell4}) are built into the
scheme regardless of the details of the $\vec \Theta$ dynamics.

Denoting the particle masses with $m_i$, their coordinates with $\vec
r_i$, and the interparticle potential (of {\em non}--electromagnetic
type) with $U$, we can thus write the Lagrangian as
\begin{eqnarray}
L & = & \sum_i \frac{m_i}{2} \dot{\vec{r_i}}^2 - U \\
\nonumber
  & + & \frac{\epsilon_0}{2c^2} 
        \int d^3 \vec r \, \dot{\vec \Theta}^2
    -   \frac{\epsilon_0}{2} \int d^3 \vec r \, \vec E^2 \\
\nonumber
  & + & \int d^3 \vec r \, \vec A \cdot \left(
        \epsilon_0 \dot{\vec E} + \vec j - 
         \epsilon_0 \vec \nabla \times \dot{\vec \Theta} \right) ;
\end{eqnarray}
here the field $\vec A$ is a Lagrange multiplier; it will later on
turn out to be the vector potential. Such a constrained variational
problem with Lagrange multipliers can be treated by Arnold's
so--called ``vakonomic'' (``variational of the axiomatic kind'')
formalism \cite{arnold}. The recipe how to do variational calculus
within that formalism is very simple: One just has to treat all
occuring variables, including the Lagrange multipliers, as if they
were independent degrees of freedom. It is thus straightforward to
obtain the equations of motion. Variation with respect to $\vec A$
just yields the fourth Maxwell equation, Eq. \ref{eq:newmaxwell4}.
Variation with respect to $\vec E$ yields
\begin{equation} \label{eq:relationEwithA}
\vec E = - \dot{\vec A} ,
\end{equation}
while from variation with respect to $\vec \Theta$ we obtain
\begin{equation} \label{eq:motionoftheta}
\frac{1}{c^2} \ddot{\vec \Theta} = \vec \nabla \times \dot{\vec A} .
\end{equation}
This is equivalent to the remaining two Maxwell equations, Eqs.
\ref{eq:maxwell2} and \ref{eq:maxwell3}: Inserting Eq.
\ref{eq:relationEwithA}, plus $\vec H = \epsilon_0 \dot{\vec \Theta}$,
into Eq. \ref{eq:motionoftheta}, we obtain directly Eq.
\ref{eq:maxwell2}. Furthermore, we can integrate Eq.
\ref{eq:motionoftheta} in time, which, together with $\vec H =
\epsilon_0 \dot{\vec \Theta}$, and the initial condition ($\vec A$ and
$\dot{\vec \Theta}$ both vanish at time $t = 0$) yields
\begin{equation} \label{eq:relationHwithA}
\vec H = \epsilon_0 c^2 \vec \nabla \times \vec A .
\end{equation}
Taking the divergence of this equation, one directly obtains Eq.
\ref{eq:maxwell3}. The interpretation of $\vec A$ as the vector
potential is also obvious from Eqs. \ref{eq:relationEwithA} and
\ref{eq:relationHwithA}, since these are the standard relations
between the electromagnetic fields and the vector potential. It should
be noted that our derivation has led us in a natural way to the
so--called temporal gauge \cite{landaulifshitz2} where the scalar
potential vanishes identically, and there is no restriction on $\vec
A$.

For deriving the equations of motion for the particles, we first note
that charge and current densities are written as
\begin{eqnarray}
\rho & = & \sum_i e_i \delta \left( \vec r - \vec r_i \right) \\
\vec j & = & \sum_i e_i \dot{\vec r_i} 
\delta \left( \vec r - \vec r_i \right)
\end{eqnarray}
where $e_i$ is the charge of the $i$th particle. Hence the current
term in the Lagrangian is written as
\begin{equation}
\int d^3 \vec r \vec A \cdot \vec j =
\sum_i e_i \vec A \left( \vec r_i \right) \cdot \dot{\vec{r_i}} .
\end{equation}
After a few lines of algebra one then finds the particle equations of
motion:
\begin{equation} \label{eq:motionofparticles1}
m_i \ddot{ \vec{r_i} } = - \frac{\partial U}{\partial \vec r_i} +
\vec F_L ,
\end{equation}
where the Lorentz force $\vec F_L$ is given by Eq.
\ref{eq:lorentzforce}.

To summarize: The requirement of local updates, combined with treating
the deviations from the BOS in the CP manner, has led us in a natural
way to standard electromagnetism, where the temporal gauge turns out
to be the most appropriate one for our purposes. It should be stressed
that this is a consistent non--relativistic setting, where the
equations of motion are valid in one particular chosen frame of
reference.

As it is common practice in electromagnetism \cite{jackson}, we can
now simplify the Lagrangian treatment by considering $\vec A$ as the
(only) field degree of freedom, while $\vec E$ and $\vec H$ are
derived quantities according to Eqs. \ref{eq:relationEwithA} and
\ref{eq:relationHwithA}. The dynamical system of charges and
electromagnetic field is then completely described by (i) the equation
of motion for the particles, Eq. \ref{eq:motionofparticles1}, and (ii)
the fourth Maxwell equation, Eq. \ref{eq:maxwell4}, which is the
inhomogeneous wave equation for $\vec A$:
\begin{equation}
\frac{\partial^2}{\partial t^2} \vec A = 
- c^2 \vec \nabla \times \left( \vec \nabla \times \vec A \right)
+ \frac{1}{\epsilon_0} \vec j .
\end{equation}
To derive these two equations of motion, it is sufficient to consider
the Lagrangian
\begin{eqnarray}
L & = & \sum_i \frac{m_i}{2} \dot{\vec{r_i}}^2 - U \\
\nonumber
  & - & \frac{\epsilon_0 c^2}{2} 
        \int d^3 \vec r \, \left( \vec \nabla \times \vec A \right)^2
    +   \frac{\epsilon_0}{2} \int d^3 \vec r \, \dot{\vec A}^2 \\
\nonumber
  & + & \int d^3 \vec r \, \vec A \cdot \vec j .
\end{eqnarray}
This dynamics has a couple of very desirable properties: Firstly,
since the dynamics is manifestly Hamiltonian (it is derived from a
Lagrangian), it conserves the phase--space volume and the energy, the
latter being given by
\begin{eqnarray} \label{eq:totalenergy}
{\cal H} & = & \sum_i \frac{m_i}{2} \dot{\vec{r}}_i^2 + U \\
\nonumber &&
+ \frac{\epsilon_0}{2} \int d^3 \vec r \, \vec E^2
+ \frac{1}{2 \epsilon_0 c^2} \int d^3 \vec r \, \vec H^2 .
\end{eqnarray}
Furthermore, one can show that the total momentum, given by
\begin{equation}
\vec P = \sum_i m_i \dot{\vec{r}}_i + \frac{1}{c^2}
\int d^3 \vec r \, \vec E \times \vec H ,
\end{equation}
is conserved as well. For the proof one can employ the dynamic
equations for the particles and fields, and make use of the identity
\begin{equation}
\int d^3 \vec r \, \vec X \times 
\left( \vec \nabla \times \vec X \right) =
\int d^3 \vec r \, \vec X \left( \vec \nabla \cdot \vec X \right) ,
\end{equation}
which holds for any vector field $\vec X$ as long as partial
integration with vanishing boundary terms can be applied.

At this point, we modify the equations of motion by discarding the
magnetic force on the particles,
\begin{equation}
m_i \ddot{\vec{r}}_i = - \frac{\partial U}{\partial \vec r_i}
+ e_i \vec E (\vec r_i) .
\end{equation}
This simplifies the algorithm significantly, while the most important
features still hold. Of course, this modified dynamics is no longer
Hamiltonian. Nevertheless, the energy, as given by Eq.
\ref{eq:totalenergy}, is still conserved.  Furthermore, the (properly
defined) phase space volume is also conserved. In order to see this,
we first write the equations of motion in pseudo--Hamiltonian style as
\begin{eqnarray} 
\label{eq:pseudohamil1}
\frac{d}{dt} \vec r_i & = & \frac{1}{m_i} \vec p_i \\
\label{eq:pseudohamil2}
\frac{d}{dt} \vec p_i & = & - \frac{\partial U}{\partial \vec r_i}
                            + e_i \vec E (\vec r_i) \\
\label{eq:pseudohamil3}
\frac{\partial}{\partial t} \vec A & = & - \vec E \\
\label{eq:pseudohamil4}
\frac{\partial}{\partial t} \vec E & = & 
c^2 \vec \nabla \times \left( \vec \nabla \times \vec A \right)
- \frac{1}{\epsilon_0} \vec j ,
\end{eqnarray}
where the $\vec p_i$ are the kinematic (and {\em not} the canonically
conjugate!) particle momenta, and the fields $\vec A$ and $\vec E$
(roughly) play the roles of coordinates and momenta, respectively.

Now, phase space volume conservation for some (high--dimensional)
dynamical system, given by the equation
\begin{equation}
\dot{\vec x} = \vec f \left( \vec x \right) ,
\end{equation}
where $\vec x$ comprises the set of all phase--space variables, holds
if and only if $\vec \nabla \cdot \vec f = 0$ (in analogy to
incompressible flow in hydrodynamics). It is trivially checked that
this relation does hold for our system.

However, momentum conservation does {\em not} hold for our modified
dynamics. The momentum carried away by the electromagnetic waves is
not completely balanced by the particle momenta.  Rather, we have the
relation
\begin{equation}
\sum_i \vec p_i = \mbox{\rm const.} + O \left( c^{-2} \right) .
\end{equation}
This is not a catastrophe, since momentum conservation is usually only
important in studies of dynamics. However, for such calculations one
has to use a fairly large value of $c$ anyways, since otherwise the
electromagnetic field is not in its quasi--static limit, and the
particle trajectories get too much distorted. Furthermore, one must
expect that momentum conservation is also violated as a result of
the lattice discretization, which breaks the translational invariance
of the system.

We now assume that the dynamics is sufficiently nonlinear to make the
system ergodic. This seems reasonable for the case of a many--charge
system, in particular if the potential $U$ has a strongly repulsive
core to facilitate ``collisions''. We therefore assume that the system
has no further important conservation law except for the fact that it
stays on the CS, and that the energy ${\cal H}$ is conserved. The
additional conserved quantities mentioned in Ref. \cite{maggs_prl_2}
probably apply only to the charge--free case, in which the system is
harmonic and hence integrable. We can hence apply standard statistical
physics to the system and assume that the dynamics results in an
equidistribution of states in terms of the variables $\vec r_i, \vec
p_i, \vec A$ and $\vec E$ (microcanonical ensemble). Making use of the
fact that thermodynamic ensembles are equivalent in the large--system
limit, we can instead employ the canonical ensemble, which is
easier. With $\beta = 1 / (k_B T)$, where $k_B$ is Boltzmann's
constant and $T$ the absolute temperature, we may therefore write the
partition function as
\begin{eqnarray}
Z & = & \int d \vec r_i \int d \vec p_i \int {\cal D} \vec A \int {\cal D}
\vec E \exp \left( - \beta {\cal H} \right) 
\nonumber \\
 && \times \delta \left( \vec \nabla \cdot \vec E - 
\frac{1}{\epsilon_0} \rho \right) ,
\end{eqnarray}
where ${\cal H}$ is given by Eq. \ref{eq:totalenergy}. It is now
straightforward to integrate out the momenta, the $\vec A$ field,
and the transversal component of the $\vec E$ field. The integration
over the longitudinal component of $\vec E$ cancels with the
delta function, such that the only remaining degrees of freedom
are the particle coordinates, for whose potential of mean force
we hence find
\begin{equation} \label{eq:potentialofmeanforce}
{\cal H}_{conf} = U + \frac{\epsilon_0}{2} \int d^3 \vec r \, \vec E^2 ;
\end{equation}
here $\vec E$ is nothing but the solution of the standard
electrostatic problem, Eqs. \ref{eq:statics1} and \ref{eq:statics2},
i.~e. the Coulomb field. Inserting this field into Eq.
\ref{eq:potentialofmeanforce}, we find the standard Coulomb
Hamiltonian,
\begin{equation} \label{eq:potentialofmeanforcecoulomb}
{\cal H}_{conf} = U + \frac{1}{2} \frac{1}{4 \pi \epsilon_0}
\int d^3 \vec r \, \int d^3 \vec r^\prime \,
\frac{ \rho( \vec r ) \rho (\vec r^\prime )}
{\left\vert \vec r - \vec r^\prime \right\vert} .
\end{equation}
This demonstrates that the particles behave statistically in the same
way as if they would directly interact Coulombically. This concludes
the derivation. On the lattice, however, we have to take into account
that the above Hamiltonian includes unphysical self--interactions,
which we have to subtract (without such a subtraction the
self--interaction would ultimately, i.~e. in the continuum limit of
vanishing lattice spacing, completely dominate the behavior), and that
instead of the $1 / r^2$ Coulomb field we have to insert the
lattice--discretized solution of the lattice equations corresponding
to Eqs. \ref{eq:statics1} and \ref{eq:statics2}. This shall be
discussed in the next section.

\section{Discretization, Lattice Green's Function and Self--Interaction}
\label{sec:selfenergy}

For implementation on the computer, the equations need to be
discretized with respect to both space and time. For the moment, we
will only consider the spatial discretization, and consider time still
as a continuous variable. The issue of time discretization is
discussed in \ref{app:integrator}.

We use a spatial discretization scheme \cite{maggs_prl_1,yee} where
the charges are interpolated linearly to the eight surrounding lattice
sites of a simple--cubic lattice. The currents, as well as the fields
$\vec E$ and $\vec A$ are put onto the connecting links. The curl of
link variables is put onto the lattice plaquettes, and the curl of
plaquette variables onto the links (in both cases one uses the four
fields which encircle the result). Furthermore the divergence of link
variables is put onto the sites, using the adjacent fields, while
the gradient of a scalar variable located on the sites is a link
variable. For more details, see \ref{app:discretization}.

Let us now discuss how the Coulomb potential looks on the
lattice. Obviously, we have to solve Eqs. \ref{eq:statics1} and
\ref{eq:statics2} on the lattice. As in the continuum, we can take
into account the longitudinal character of the electric field by the
ansatz
\begin{equation}
\vec E = - \vec \nabla \phi ,
\end{equation}
where $\phi$ is the electrostatic potential on the sites, and
$\vec \nabla$ is the lattice--discretized gradient. We thus obtain
the Poisson equation on the lattice,
\begin{equation}
- \vec \nabla^2 \phi = \frac{1}{\epsilon_0} \rho ,
\end{equation}
where the lattice version of the operator $\vec \nabla^2$ is clear from
the previous definitions of gradient and divergence.

A system with periodic boundary conditions is invariant with
respect to lattice translations, and this allows us to write
\begin{equation}
\phi (\vec r) = \frac{a^2}{\epsilon_0} \sum_{\vec r^\prime}
G(\vec r - \vec r^\prime) \rho (\vec r^\prime) ,
\end{equation}
where $a$ is the lattice spacing, $\vec r$ denotes the sites, and $G$
is the lattice Green's function, obeying the equation
\begin{equation}
- \vec \nabla^2 G(\vec r) = \frac{1}{a^2} \delta (\vec r) ,
\end{equation}
where $\delta (\vec r)$ is the Kronecker symbol, i.~e.
$\delta (\vec r) = 1$ for $\vec r = 0$, and $\delta (\vec r) = 0$
for all other lattice sites.

For an $L_x \times L_y \times L_z$ lattice with periodic boundary
conditions the solution can be obtained straightforwardly via discrete
Fourier transformation. At the site $\vec r = a (n_x, n_y, n_z)$ one
finds
\begin{eqnarray}
\nonumber
G \left( \vec r \right)
& = & 
{\sum_{p_x = 0}^{L_x - 1}}^\prime \exp 
\left( 2 \pi i \frac{p_x n_x}{L_x} \right)  \\
\nonumber
& \times &
{\sum_{p_y = 0}^{L_y - 1}}^\prime \exp 
\left( 2 \pi i \frac{p_y n_y}{L_y} \right) \\
\nonumber
& \times &
{\sum_{p_z = 0}^{L_z - 1}}^\prime \exp 
\left( 2 \pi i \frac{p_z n_z}{L_z} \right) \\
& \times & \frac{1}{L_x L_y L_z}
\tilde G \left( p_x, p_y, p_z \right) ,
\end{eqnarray}
where $\sum^\prime$ indicates that $(p_x, p_y, p_z) = (0,0,0)$ is
excluded (for reasons of overall charge neutrality), and $\tilde G$ is
given by
\begin{eqnarray}
\nonumber
\tilde G \left( p_x, p_y, p_z \right)^{-1} = 6
& - & 2 \cos \left( 2 \pi \frac{p_x}{L_x} \right) \\
& - & 2 \cos \left( 2 \pi \frac{p_y}{L_y} \right) \\
\nonumber
& - & 2 \cos \left( 2 \pi \frac{p_z}{L_z} \right) .
\end{eqnarray}
A lot is known about this function, in particular in the limit $L_i
\to \infty$ \cite{katsura1,katsura2,glasser}. For our purposes,
however, it is sufficient to note that (i) $G$ can be calculated at
the beginning of the simulation once and for all, including the finite
size effect, and that (ii) $G ( \vec r = 0 )$ is {\em finite}, even in
the limit $L_i \to \infty$ (but, of course, keeping $a$ fixed).

We thus find for the potential of mean force (cf. Eq.
\ref{eq:potentialofmeanforcecoulomb})
\begin{equation} \label{eq:potentialofmeanforcediscrete}
{\cal H}_{conf} = U + \frac{1}{2} \frac{1}{\epsilon_0 a}
\sum_{\vec r} \sum_{\vec r^\prime} G(\vec r - \vec r^\prime)
q (\vec r) q (\vec r^\prime) ,
\end{equation}
where $q(\vec r) = a^3 \rho (\vec r)$ is the charge on site $\vec r$.
Now, the charges on the sites are related to the charges $e_i$ on the
particles via the interpolation scheme, $q(\vec r) = \sum_i e_i s(\vec
r, \vec r_i)$ and $q(\vec r^\prime) = \sum_j e_j s(\vec r^\prime, \vec
r_j)$, where $s$ is the ``smearing'' function. Inserting this into
Eq. \ref{eq:potentialofmeanforcediscrete}, we find an effective
interaction between different particles $i \ne j$, but also an
unphysical self--energy term for $i = j$. This is given by
\begin{equation}
U_{self,i} = \frac{1}{2} \frac{1}{\epsilon_0 a}
\sum_{\vec r} \sum_{\vec r^\prime} G(\vec r - \vec r^\prime)
e_i^2 s(\vec r, \vec r_i) s(\vec r^\prime, \vec r_i) .
\end{equation}
This depends explicitly on the particle coordinate $\vec r_i$. The
physical interpretation is simply that the Coulomb repulsion from the
interpolated charges on the cube corners tries to drive the particle
into the center of the cube. For small lattice spacings, this effect
dominates over all other interactions, and therefore must be taken
care of. For this reason, we add a term $- \sum_i U_{self,i}$ to the
interparticle potential $U$, and apply the corresponding force to the
particles. This is feasible since $G$ is known explicitly, and the
smearing function $s$ is short--ranged, such that $\sum_{\vec r}$ runs
over eight sites only.

\section{Yukawa Subtraction}
\label{sec:yukawa}

Rottler and Maggs \cite{maggs_prl_2} suggest another subtraction
scheme which has the nice property of introducing another optimization
parameter $\kappa$ into the method. Essentially, interactions up to
the length scale $\kappa^{-1}$ are done in real space, while only the
residual long--range part beyond $\kappa^{-1}$ is treated via the
dynamics. The disadvantage, however, is that it does not treat the
lattice effects completely rigorously. We hence believe that probably
the best method consists of a combination between our lattice Green's
function subtraction, and their ``dynamic Yukawa'' approach.

In order to understand the latter, let us first consider the
functional
\begin{equation}
{\cal F} = - \frac{\epsilon_0}{2} \int d^3 \vec r \,
\left( \vec \nabla \phi \right)^2 +
\int d^3 \vec r \, \rho \phi
\end{equation}
and study, for fixed $\rho$,
\begin{equation}
\frac{\delta {\cal F}}{\delta \phi} = 0 .
\end{equation}
It is straightforward to see that (i) this variational problem
is equivalent to the Poisson equation for the electrostatic
potential $\phi$, and that (ii) insertion of the solution
into ${\cal F}$ yields ${\cal F} = + (1/2) \int d^3 \vec r \,
\rho \phi$, i.~e. the correct electrostatic energy. However,
this functional is useless for dynamic simulations where one would try
to simulate a coupled dynamics of $\rho$ and $\phi$. The reason is
that the $\vec \nabla \phi$ term has the wrong sign, such that
arbitrarily large variations of $\phi$ are favored and the simulation
would be inherently unstable (the partition function for integrating
out the $\phi$ field would not exist).

A well--behaved theory, however, is obtained by just turning
the sign of the $\vec \nabla \phi$ term:
\begin{equation}
{\cal F} = + \frac{\epsilon_0}{2} \int d^3 \vec r \,
\left( \vec \nabla \phi \right)^2 +
\int d^3 \vec r \, \rho \phi .
\end{equation}
This results in $+ \vec \nabla^2 \phi = \rho / \epsilon_0$,
and insertion into the functional yields again
${\cal F} = + (1/2) \int d^3 \vec r \, \rho \phi$.
Since, however, $\phi$ is just the negative of the
real (physical) electrostatic potential, one obtains
a theory which describes attraction between like charges
and repulsion between unlike charges. We now introduce
an additional field degree of freedom $\phi$, and couple
this to the original method (Lagrangian) via
\begin{eqnarray}
L & \to & L + \frac{\epsilon_0}{2 c_\phi^2} 
\int d^3 \vec r \, \dot{\phi}^2 \\
\nonumber
&&
- \frac{\epsilon_0}{2} \int d^3 \vec r \,
\left( \vec \nabla \phi \right)^2 -
\int d^3 \vec r \, \rho \phi .
\end{eqnarray}
Here $c_\phi$ is another dynamical parameter of dimension velocity. It
can be set identical to $c$, but need not.  This modified method would
result in an additional potential of mean force between the charges
which would {\em exactly cancel} the original Coulomb potential
(including self--terms). This is apparently not useful. However, we
can introduce a slightly modified coupling with a screening parameter
$\kappa > 0$:
\begin{eqnarray}
L & \to & L + \frac{\epsilon_0}{2 c_\phi^2} 
\int d^3 \vec r \, \dot{\phi}^2 \\
\nonumber
&&
- \frac{\epsilon_0}{2} \int d^3 \vec r \,
\left( \vec \nabla \phi \right)^2
- \frac{\epsilon_0}{2} \int d^3 \vec r \,
\kappa^2 \phi^2
- \int d^3 \vec r \, \rho \phi .
\end{eqnarray}
This introduces an additional potential of mean force between
the charges, which, in the continuum limit, would read
\begin{equation}
U_Y(r) = - \frac{1}{4 \pi \epsilon_0} \frac{e_i e_j}{r} \exp(- \kappa r) ,
\end{equation}
such that unlike charges repel each other with a screened Coulomb
interaction. This weakens the original Coulomb interactions on a local
scale, and can be corrected by adding $- U_Y$ to the standard
interparticle potential. Here one can use the continuum version of the
potential; this will only serve to decrease the influence of lattice
artifacts.

In principle, this also alleviates the self--energy problem. However,
the lattice Green's functions of the unscreened and screened Coulomb
case are slightly different and only coincide in the limit $\kappa \to
0$. In preliminary tests we found that fairly small screening
parameters are needed to overcome the self--energy problem with high
accuracy. We therefore believe that one should rather try to subtract
the self--energy for both the unscreened and the screened interaction
separately by the respective exact lattice Green's function. In this
case, the Yukawa subtraction would no longer serve the purpose of
overcoming self--energies, but rather to resolve interparticle
interactions rather faithfully on a local scale, such that (hopefully)
larger lattice spacings are feasible. Further investigations are
necessary on this issue.

\section{Numerical Results}
\label{sec:results}

As a simple test system, we have studied $N$ charged particles
in a cubic box with periodic boundary conditions. They interact
via a purely repulsive Lennard--Jones (LJ) potential
\begin{equation} \label{eq:replj}
  U_{LJ} = \left\{%
      \begin{array}{l l}
        4\epsilon \left[ \left( \displaystyle\frac{\sigma}{r} \right)^{12}
          -\left( \displaystyle\frac{\sigma}{r} \right)^6
          +\displaystyle\frac{1}{4} \right]
          & \hspace{0.5cm} r \le 2^{1/6}\sigma \\
        0 & \hspace{0.5cm} r \ge 2^{1/6}\sigma
      \end{array}
\right.  .
\end{equation} 
We choose a unit system where the potential parameters $\sigma$ and
$\epsilon$, as well as the particle mass $m$, are set to unity. Time
is thus measured in units of $\tau_{LJ} = \sqrt{m \sigma^2 /
\epsilon}$. We study systems at temperature $k_B T = 1$
and particle number density $\rho = 0.07$. The equations of motion
were integrated by the algorithm outlined in \ref{app:integrator}
(no Yukawa subtraction), using a time step $h = 0.01$. The friction
constant for the Langevin thermostat was set to $\zeta = 1$.

Each particle is assigned a charge $\pm e$. The strength of the
electrostatic interaction is given in terms of the Bjerrum length
\begin{equation} \label{eq:bjerrum}
l_B = \frac{e^2}{4 \pi \epsilon_0 k_B T} .
\end{equation}
We first started out with the value $l_B = 20$ (rather strong
electrostatic coupling). We chose this system because it had been
studied previously by P$^3$M \cite{espressowebsite}. However, it has
turned out that this is not the best state point for a benchmark,
since the coupling is so strong that it actually induces phase
separation (gas--liquid transition). This is in accord with the phase
diagram presented in Ref. \cite{depablo}; the system studied there is
not too different from ours.

\begin{figure}
\centerline{
\includegraphics[width=10cm]{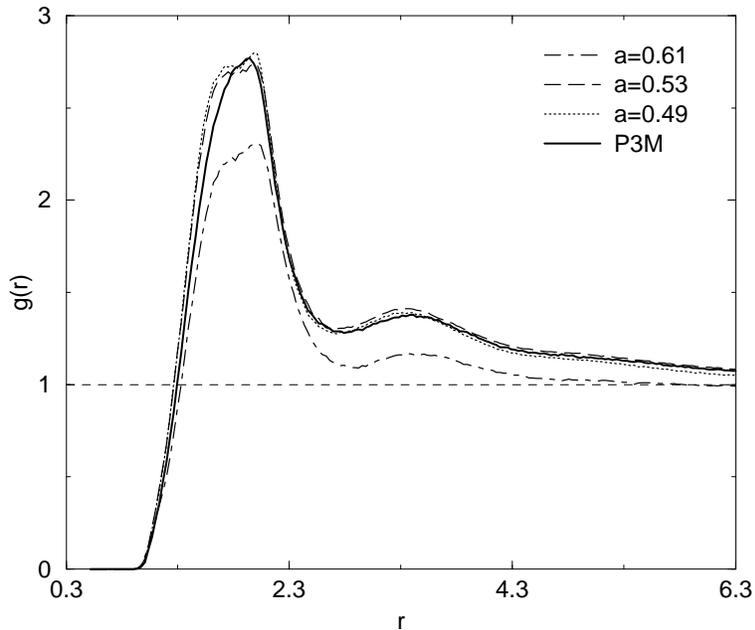}
}
\caption{Pair correlation function of like charges
at density $\rho = 0.07$ and Bjerrum length $l_B = 20$,
comparing data obtained with P$^3$M with those from
MEMD for different lattice spacings.}
\label{fig:corrplusplus20}
\end{figure}

\begin{figure}
\centerline{
\includegraphics[width=10cm]{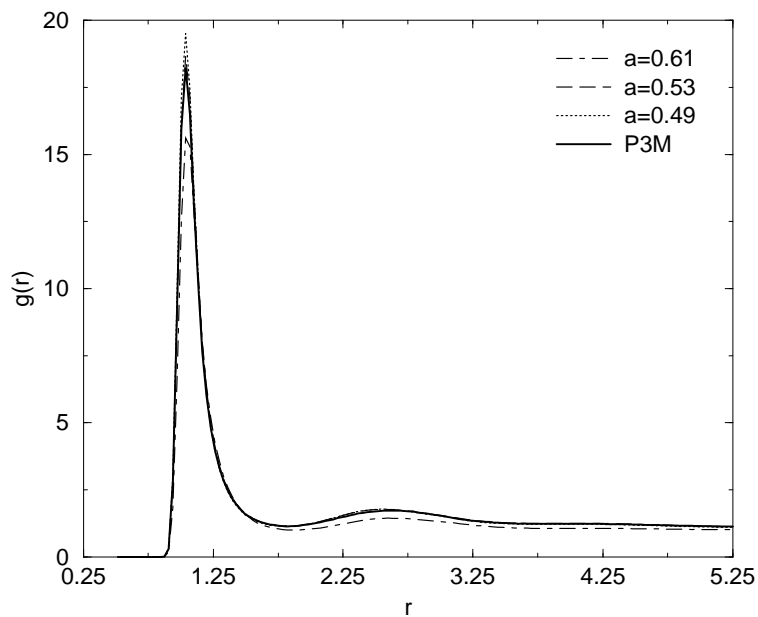}
}
\caption{Pair correlation function of unlike charges
at density $\rho = 0.07$ and Bjerrum length $l_B = 20$,
again comparing P$^3$M with MEMD.}
\label{fig:corrplusminus20}
\end{figure}

The number of particles was set to $N = 2000$. Both the P$^3$M and the
MEMD calculations were done within the framework of the ``ESPResSo''
software package \cite{espressowebsite} of the Theory Group at the MPI
for Polymer Research, Mainz. In both cases we used a program version
which was fully parallelized, based upon domain decomposition. The
P$^3$M parameters were optimized using an automatized routine building
upon the work of Refs. \cite{deserno1,deserno2}, where a formula for
the relative error of the force per particle, $\Delta F / F$, was
derived. The routine provides optimized simulation parameters after an
upper bound for $\Delta F / F$ has been supplied. For our system, we
required an accuracy of $10^{-3}$, resulting in the following
parameters: Mesh size $32^3$; 5th order charge assignment; real--space
cutoff $8.2$; $\alpha = 0.36$ (this parameter controls the split--up of
the computational load between real and Fourier space). For the MEMD
calculations, we used $c = 1$ and varied the lattice spacing $a$.

The pair correlation functions of this system are shown in
Figs. \ref{fig:corrplusplus20} and \ref{fig:corrplusminus20}. The runs
were long enough to equilibrate the system reasonably well on the
local scale. As a control, we also ran a more accurate P$^3$M
simulation and found no visible difference from the original P$^3$M
result. The MEMD results in turn confirm that the static correlations
converge towards those of the real electrostatic system when the
lattice spacing $a$ decreases. We found a value of $a = 0.53$
acceptable, corresponding to a $58^3$ lattice for the $N = 2000$
system. For such a fine lattice, there is practically never more than
one particle per cube.

We did not study this system further, since benchmarks at this state
point are severely hampered by the gas--liquid transition: On the one
hand, the system needs a long time to equilibrate (i.~e. to condense
macroscopically), and on the other hand the particle density is very
inhomogeneous in the relaxed state. This, in turn, is very detrimental
to efficient geometric parallelization, since some processors have to
treat very many particles, while others work on essentially none. In
other words, such a system will behave very poorly with respect to
load--balancing, and will give no good hint on the parallelization
efficiency under normal (homogeneous) conditions.

\begin{figure}
\centerline{
\includegraphics[width=10cm]{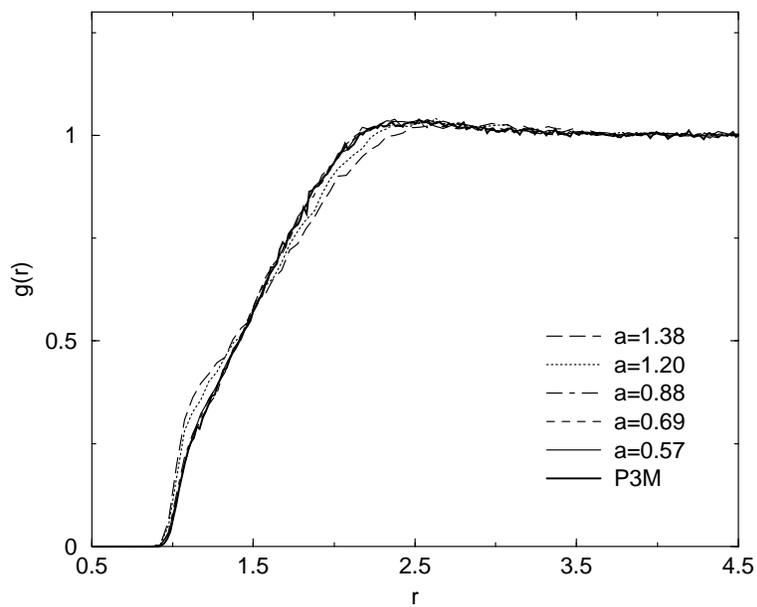}
}
\caption{Pair correlation function of like charges
at density $\rho = 0.07$ and Bjerrum length $l_B = 5$,
comparing data obtained with P$^3$M with those from
MEMD for different lattice spacings.}
\label{fig:corrplusplus5}
\end{figure}

\begin{figure}
\centerline{
\includegraphics[width=10cm]{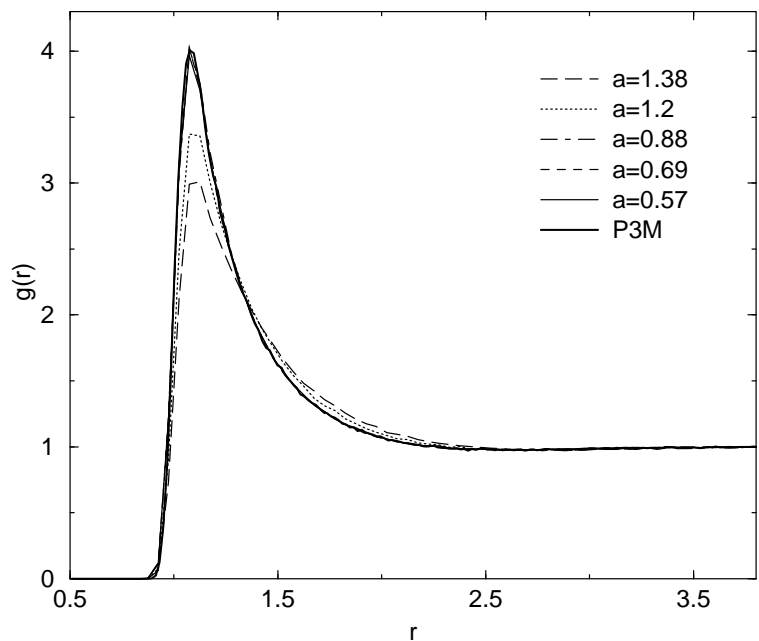}
}
\caption{Pair correlation function of unlike charges
at density $\rho = 0.07$ and Bjerrum length $l_B = 5$,
again comparing P$^3$M with MEMD.}
\label{fig:corrplusminus5}
\end{figure}

We hence abandoned this state point and instead systematically studied
the weaker coupling $l_B = 5$, at first restricting the particle
number to $N = 500$. Furthermore, we slightly increased the particle
friction coefficient to $\zeta = 1.5$. All other parameters (density,
temperature) were left unchanged. This is well in the homogeneous
phase, and the pair correlation functions show much less structure,
see Figs. \ref{fig:corrplusplus5} and \ref{fig:corrplusminus5}.  It
also turns out that here a larger MEMD lattice spacing $a = 0.88$ is
sufficient to reasonably approximate the P$^3$M result.

\begin{figure}
\centerline{
\includegraphics[width=10cm]{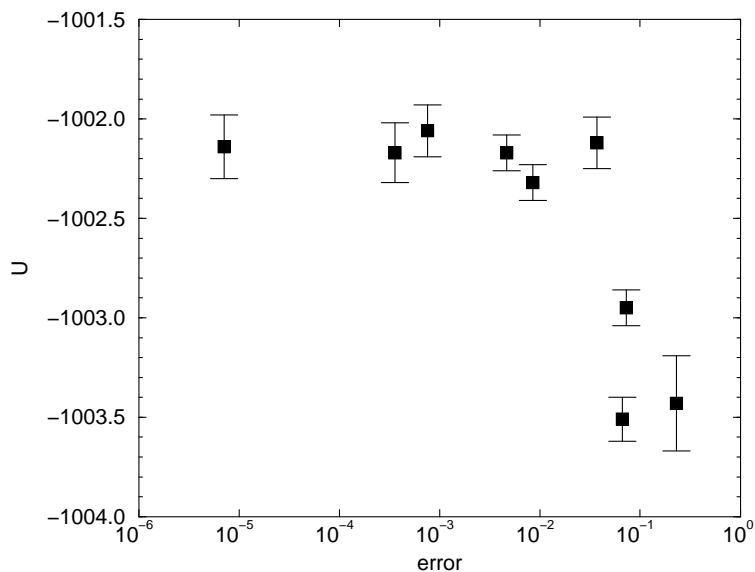}
}
\caption{Internal electrostatic energy for P$^3$M,
as a function of the accuracy parameter $\Delta F / F$.}
\label{fig:calibrationp3m}
\end{figure}

Our next aim is to compare the efficiency of P$^3$M and MEMD. In order
to do this in a meaningful way, it is necessary to make sure that (i)
both methods use parameters which yield roughly the same accuracy in
the representation of the electrostatic interaction, and that (ii)
both methods use parameters for which the results are obtained most
quickly, within the accuracy constraint. The (thermally averaged)
electrostatic energy $U$ is a variable which, on the one hand, is easy
to evaluate, and, on the other hand, reasonably sensitive to the
long--range correlations between the particles. We therefore used this
observable for calibrating the accuracy of the simulations. For
P$^3$M, we therefore calculated $U$ as a function of the accuracy
parameter $\Delta F / F$. The results are shown in Fig.
\ref{fig:calibrationp3m}. The error bars were obtained as statistical
error bars, using the block average method \cite{flyvbjerg}. From
these results, one sees that an accuracy parameter of $\Delta F / F =
3.7 \times 10^{-2}$ is good enough. This corresponds to the following
P$^3$M parameters: Mesh size $16^3$, third--order charge assignment,
real--space cutoff $4.4$, $\alpha = 0.43$.

\begin{figure}
\centerline{
\includegraphics[width=10cm]{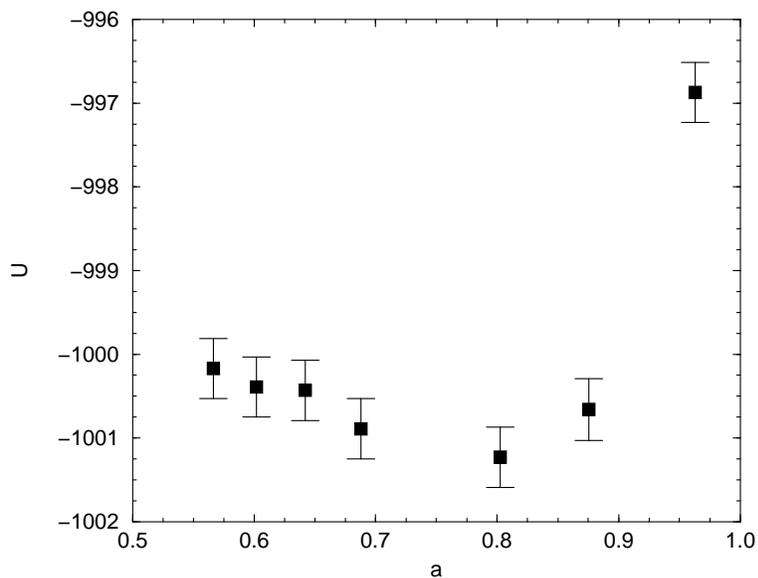}
}
\caption{Internal electrostatic energy for MEMD,
as a function of the lattice spacing $a$.}
\label{fig:calibrationmemd}
\end{figure}

At this point, it is necessary to comment on the evaluation of $U$ in
MEMD. The electric and magnetic field energies are given by $ (
\epsilon_0 / 2 ) \int d^3 \vec r \, \vec E^2 $ and $ ( 1 / (2
\epsilon_0 c^2 ) ) \int d^3 \vec r \, \vec H^2 $, respectively. Both
types of fields have one longitudinal and two transversal degrees of
freedom per lattice site. The longitudinal magnetic degrees of freedom
are however not excited, due to $\vec \nabla \cdot \vec H =
0$. Furthermore, the magnetic part of the Hamiltonian is strictly
quadratic, and the equipartition theorem can be applied. For this
reason, the thermally averaged magnetic field energy is just given by
$M k_B T$, where $M$ is the number of lattice sites. We have checked
this relation, and found good agreement, except for a deviation of a
few percent, which decreases with the time step, and must hence be
attributed to discretization errors --- the exact Boltzmann
distribution is only generated in the limit of vanishing time
step. This finding is a strong support of our belief that, except for
the obvious conservation laws for the longitudinal fields, there are
no further hidden conserved quantities in the system, and a
thermostatting of the magnetic field is not necessary. We can also
apply the equipartition theorem to the transversal part of the
electric field energy, and hence the thermally averaged Coulomb
energy is given by
\begin{equation}
U = \frac{\epsilon_0}{2} \int d^3 \vec r \, \left< \vec E^2 \right>
  - M k_B T - \left< U_{self} \right> ,
\end{equation}
where $U_{self}$ is the self--energy, as discussed in Sec.
\ref{sec:selfenergy}. We have measured $U$ as a function of the
lattice spacing $a$, using this recipe. While the results were in
reasonable agreement with the P$^3$M results, we did not find
convergence for $a \to 0$. Rather, it seems that $U$ diverges for
small $a$ (apparently like $1 / a$, though the data are not precise
enough to be sure). It turns out that this divergence is reduced by
reducing the time step, i.~e. it is again an effect of discretization
errors. Our explanation is that the cancellation of the self--energy
is not perfect, because the subtraction term assumes the exact Coulomb
lattice Green's function, while the simulation produces an effective
lattice Green's function, which is slightly distorted by
discretization errors. This effect is crucial for the energy
calculation (and probably also for the evaluation of the pressure, and
related quantities), but not for the particle configurations, which
are stabilized by the repulsive LJ interactions. Probably that problem
must be solved by combining MEMD with a Monte Carlo procedure, which
can enforce strict detailed balance, and thus produce the {\em exact}
Boltzmann distribution. For the moment, we just solved the problem by
taking the particle configurations produced by MEMD, and using
accurate P$^3$M for evaluating $U$.  The results, which now do
converge, and give good agreement with the P$^3$M data, are shown in
Fig. \ref{fig:calibrationmemd}. Again, we attribute the small
remaining systematic deviation to discretization errors. Taking all
these considerations together, we took the lattice constant $a = 0.88$
as a value which produces sufficiently accurate data, consistent with
our findings from the pair correlation function.

\begin{figure}
\centerline{
\includegraphics[width=10cm]{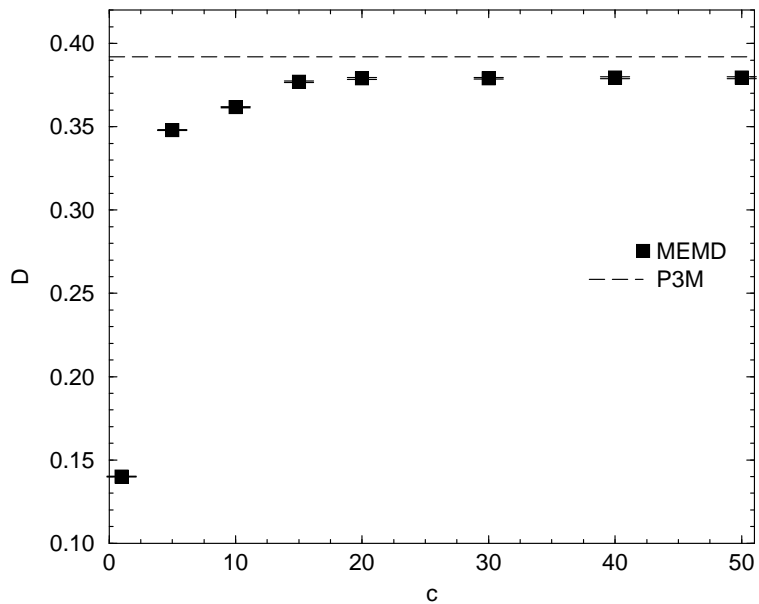}
}
\caption{Particle diffusion constant as a function of the
speed of light $c$.}
\label{fig:diffusion}
\end{figure}

Having fixed the parameters for accuracy, we now turn to optimizing
the speed. In P$^3$M, this was already done by the automatic routine
(see above). In MEMD, we still have the speed of light $c$ at our
disposal. This parameter only influences the dynamics of the system,
but not the statics. We actually checked that $U$ does not
systematiclly vary with $c$ within our error bars --- such a
dependence could still be possible as a result of discretization
errors. Furthermore, the CPU time necessary for one update step does
not depend on $c$. Hence one would like to take a value of $c$ for
which the configurations decorrelate particularly quickly. In
principle, each observable and its time autocorrelation function would
have to be considered separately \cite{flyvbjerg}. This is of course
impractical, and hence we have taken the simpler criterion that the
diffusion constant $D$ of the particles, obtained from their mean
square displacement, should be maximized. The corresponding data are
shown in Fig. \ref{fig:diffusion}. One sees that $D$ first increases
as a function of $c$, but then saturates at a value which is in good
agreement with the P$^3$M value, except for some discretization
errors. This nicely confirms the expectation that the dynamic
properties should converge to electrostatic behavior for $c \to
\infty$.  We thus have the very favorable situation that computational
efficiency for the statics and reasonable reproduction of the dynamics
are not mutually exclusive, but rather congruent. However, this does
not mean that one should just simply take a huge $c$ value. Rather,
$c$ has to be small enough such that the Courant stability criterion
\cite{numrecip}, $c \ll a / h$, is still satisfied.  For our
parameters, $a / h = 88$, and hence $c$ should be significantly
smaller (in fact, our program crashed for $c = 55$). Therefore, we use
the value $c = 20$.

For the calibrated and optimized parameters, it makes sense to look at
the speed in terms of CPU time. We ran the system on an AMD Athlon MP
2000+ processor for 2000 MD time steps. Since the diffusion constants
for both methods are essentially identical, we do not need to take
into account different rates of decorrelation. For P$^3$M, the run used
17 seconds of CPU time, while for MEMD 16 seconds were needed. This
shows that MEMD for a system of this density is a comptetitive
alternative to P$^3$M.

\begin{figure}
\centerline{
\includegraphics[width=10cm]{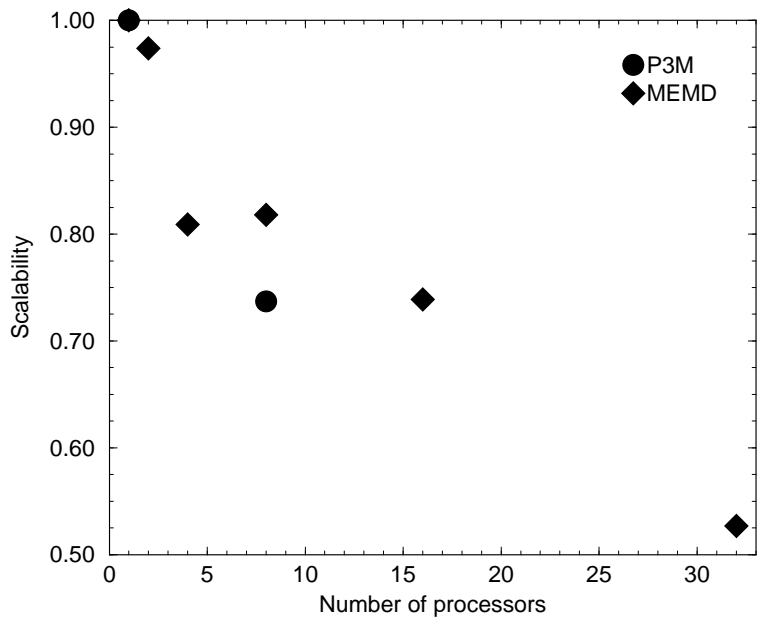}
}
\caption{Scalability factors $s$ as a function
of the number of processors, for both NEMD and P$^3$M.
For further details, see text.} 
\label{fig:scalability}
\end{figure}

Furthermore, we studied the scalability of our parallel programs at
this state point. To this end, we systematically increased the
particle number and the number of processors such that each processor
keeps $N = 4000$ particles on average. The simulations were run on an
IBM Regatta H server, where $10^4$ time steps were used for
equilibration, and another $10^4$ steps for measuring the CPU time.
The P$^3$M parameters (lattice constant of the mesh, real--space
cutoff, $\alpha$) were left unchanged throughout, since (i) one should
expect that these values are reasonably close to the optimum also for
larger systems, according to Refs.  \cite{deserno1,deserno2}, and (ii)
the automated optimization routine does not work very well for a very
large number of particles. For a single processor, the timings were
$759$ seconds for P$^3$M and $398$ seconds for MEMD. We do not know
for sure why the relative efficiency is so much different compared to
the AMD processor, but speculate this might have to do with more
efficient handling of memory for the Regatta architecture (note that
MEMD is quite memory--intensive, due to all the various field
variables). Figure \ref{fig:scalability} presents the scalabity
factors as a function of the number of processors, for both methods.
The scalability factor $s$ is defined as
\begin{equation}
s = \frac{ N_p \tau (N) }{ \tau (N_p N) },
\end{equation}
where $N$ is the number of particles on a single processor, $N_p$ the
number of processors, and $\tau$ the total CPU time for a given number
of steps. Note that for P$^3$M we can produce reasonable data only for
a processor number which is a power of eight, due to the Fast Fourier
Transform in each spatial direction. For eight processors, we find
that the scalability of MEMD is slightly better than for P$^3$M.
Furthermore, for MEMD we find reasonably acceptable (though not
excellent) scalability behavior up to $32$ processors.

\section{Conclusions}
\label{sec:conclus}

MEMD is rather easy to implement and to parallelize. The numerical
results, though being far from conclusive yet, seem to indicate that
the algorithm is a competitive alternative to existing schemes for
sufficiently dense systems. However, in electrostatic problems one
often goes to much smaller densities. If we would apply the present
MEMD method to such a dilute system, the number of grid points would
become overwhelmingly large. P$^3$M does not have this problem; due to
the split--up of the work between real space and Fourier space it is
possible to keep the number of grid points reasonably small. It is
therefore clear that MEMD for such systems can only be competitive if
it is also possible to use a reasonably coarse grid. We believe that
this might be possible by introducing Yukawa subtraction combined with
our Green's function subtraction for both the unscreened and the
screened interaction. This further optimization of the method is left
for future investigation. Another problem which needs to be addressed
is the consistent handling of the discretization errors in calculating
the energy, and related quantities --- as we have seen, these interact
in a very unfavorable way with the self--energy problems. We believe
that this can be solved by combining MEMD with Monte Carlo, such that
the Boltzmann distribution is reproduced exactly, and the potential of
mean force is known exactly. Moreover, the dynamic properties of the
algorithm have to be studied in more detail. In particular, it is
necessary to investigate the accuracy of momentum conservation, and
how this depends on the lattice spacing and the speed of light. This
latter question is particularly important when considering
applications which aim at dynamic properties, like, e.~g.  the dynamic
behavior of charged colloidal suspensions. Much remains to be done,
but the existing results are reasonably encouraging.

% \begin{acknowledgments}
\ack
Stimulating discussions with Kurt Kremer, Christian Holm, J\"org
Rottler, and, in particular, Vladimir Lobaskin and Tony Maggs are
gratefully acknowledged. We thank Bernward Mann for supplying
some of the P$^3$M benchmark data.
% \end{acknowledgments}

\begin{appendix}

\section{Details about Discretization}
\label{app:discretization}

A particularly useful spatial discretization scheme
\cite{maggs_prl_1,yee} works as follows: The charges are interpolated
onto the vertices $\vec r$ of a simple--cubic lattice with lattice
spacing $a$. If the charge $e_i$ is located at position $\vec r_i$ in
continuous space, then some nearby sites $\vec r$ are assigned some
partial charges $q_i (\vec r) = e_i s(\vec r, \vec r_i)$ ($s$ denoting
a ``smearing'' function) such that $\sum_{\vec r} q_i (\vec r) = e_i$
or $\sum_{\vec r} s(\vec r, \vec r_i) = 1$. The total charge on site
$\vec r$ is the sum of the contributions from all particles, $q(\vec
r) = \sum_i q_i (\vec r)$, and the charge density is written as $\rho
(\vec r) = a^{-3} q(\vec r)$. Different choices for $s$ are possible;
we have chosen linear interpolation to the eight vertices which form
the cube in which the particle resides:

\begin{eqnarray}
\nonumber
s(\vec r, \vec r_i) 
& = & 
\left( 1 - \frac{1}{a} \left\vert x - x_i \right\vert \right) \\
& \times &
\left( 1 - \frac{1}{a} \left\vert y - y_i \right\vert \right) \\
\nonumber
& \times &
\left( 1 - \frac{1}{a} \left\vert z - z_i \right\vert \right) ;
\end{eqnarray}
here $x$, $y$ and $z$ denote the lattice coordinates of the vertices.

Now, the vector fields $\vec j$, $\vec A$ and $\vec E$ are put on the
{\em links} which connect the vertices, in such a way that they are
aligned with the links. For instance, a link oriented along the $x$
axis would contain a variable $E$ which is the electric field at the
position of the link, and which is positive if $\vec E$ points into
the $+x$ direction, while it is negative if it points in the $-x$
direction. The divergence of such fields is put onto the sites, such
that one just takes the differences of the field values associated
with those six links which are directly connected to the site.
Conversely, the curl of such fields is put onto the {\em plaquettes}
by taking differences from the four field values which encircle the
plaquette. The result is a vector perpendicular to the plaquette;
positive (negative) field values are associated with a vector pointing
in the $+x$ ($-x$) direction (for the case that the plaquette is
perpendicular to the $x$ axis). Obviously, the fields $\vec \Theta$
and $\vec H$ must be such plaquette variables. The curl of plaquette
fields is put onto the links, by taking differences from the four
plaquettes adjacent to that link. Finally, the divergence of a
plaquette field is put into the center of the cubes, by taking
differences from the six plaquettes which enclose the cube. With these
definitions it is easy to see that the divergence of a curl vanishes
identically, as in the continuum, both for link and for plaquette
fields. Furthermore, we can define the gradient of a scalar field, the
latter being on the sites. The result is put on the links and obtained
by just taking the difference between the field values on the adjacent
sites. With this definition, one sees that the curl of a gradient
vanishes, too. These identities are extremely important, since they
allow us to decompose fields uniquely into longitudinal and
transversal components, and to apply standard procedures of vectorial
calculus also on the lattice.

The particle motion generates currents on the surrounding links.
We again use a linear interpolation scheme for $\vec j$, where
the current is distributed onto the twelve links which surround
the cube in which the particle is: For a link $l$ oriented in the
$x$ direction the current contribution from particle $i$ is
\begin{equation} \label{eq:interpolationofcurrent}
j (l) = a^{-3} e_i v_{ix} \left( w_1 (l) + w_2 (l) \right)
\end{equation}
where $v_{ix}$ is the $x$ component of the particle's velocity, while
$w_1$ and $w_2$ are the charge weight factors of the two sites which
are connected by the link. For the $y$ and $z$ direction, the
analogous procedure is applied. It is easy to see that the
space--discretized continuity equation holds exactly. Similarly, the
discretized version of the fourth Maxwell equation
(Eq. \ref{eq:maxwell4}) implies that the discretized version of Gauss'
law (Eq. \ref{eq:maxwell1}) holds exactly for all times if it holds at
time $t = 0$. The discretization scheme is therefore suitable to keep
the system on the constraint surface.

Apart from interpolating the charges and currents onto the lattice, we
also need to interpolate the electric field onto the particles in
order to calculate the electric force. Here we use the same scheme as
for the current (Eq. \ref{eq:interpolationofcurrent}), i.~e.  the
field in, e.~g., the $x$ direction is obtained by summing the fields
from the four surrounding links in $x$ direction, weighted by the sum
of the two charge weight factors of the sites connected by that link.

Using this scheme, the whole theory of Sec. \ref{sec:theory} is
consistently discretized. The system is initialized by putting
particles into the simulation cell (which has periodic boundary
conditions), assigning velocities and charges to them (of course, the
overall system is neutral), and calculating the electrostatic electric
field as follows: First, we exploit the fact that the solution of
Eq. \ref{eq:maxwell1} is trivial in one dimension. This allows us to
find a simple solution by just treating the spatial dimensions
recursively: First, the field in $z$ direction is calculated by taking
into account the differences between the mean charges of planes
perpendicular to $z$. Within the planes, we then take into account
charge differences between lines (after subtracting the average plane
charge) to obtain the field in $y$ direction. Finally the field in $x$
direction is obtained from the charge differences on the sites within
a line. This solution of course violates $\vec \nabla \times
\vec E = 0$. In order to bring the system onto the BOS, we iteratively
relax the $\vec \Theta$ field on the plaquettes until the electrostatic
field energy is minimal (cf. Eqs. \ref{eq:electricfieldenergy} --
\ref{eq:solutionelectrostaticminimum}). For a single plaquette, this
can be done in a single step. For the overall system, we use a
checkerboard decomposition which allows easy parallelization.  The
field $\vec A$ is initialized as zero. Then
Eqs. \ref{eq:pseudohamil1}--\ref{eq:pseudohamil4} are integrated in
time.

\section{Integrator}
\label{app:integrator}

Ideally, one would like to run MEMD via an integrator which leaves the
phase--space volume invariant and is time--reversible, such as the
Verlet algorithm in standard MD \cite{frenkelsmit}. Since the
equations of motion (even in the lattice--discretized case) have these
properties, it is indeed possible to construct such a
scheme. Disregarding Yukawa subtraction for the time being, an analog
to the Verlet algorithm for MEMD would be the following integrator for
Eqs. \ref{eq:pseudohamil1}--\ref{eq:pseudohamil4}, based upon a time
step $h$:

\begin{enumerate}
\item Update the particle momenta by half a time step.
\item Update the $\vec A$ field by half a time step.
\item Update the particle positions by half a time step.
\item Update the electric field by a full time step.
\item Update the particle positions by half a time step.
\item Update the $\vec A$ field by half a time step.
\item Update the particle momenta by half a time step.
\end{enumerate}

Here, ``update'' means the simple Euler rule $x(t + h) = x(t) +
\dot{x} (t) h$. The time consuming part (update of the particle
momenta, update of the electric field) is arranged in such a way that
only one ``force calculation'' per time step is necessary.  This
scheme does conserve the phase--space volume and is time--reversible,
however, it suffers a severe disadvantage: The update of the electric
field (step 4) is based upon a particle configuration (in real space
and velocity space) which has so far only progressed by half a time
step. As a consequence, Gauss' law is not satisfied within machine
accuracy, but rather only within the accuracy of the time
discretization (to satisfy it exactly would require to know the
current at the end of the time step, too). This is very undesirable,
and hence we have adopted the elegant solution which was found by
Rottler and Maggs \cite{maggs_prl_2} and allows to conserve {\em both}
time--reversibility and phase--space volume conservation, while
keeping the system strictly on the CS:

\begin{enumerate}
\item Update the particle momenta by half a time step.
\item Update the $\vec A$ field by half a time step.
\item Update the particle positions in $x$ direction by half a time step.
\item Update the electric field in $x$ direction by half a time step.
\item Update the particle positions in $y$ direction by half a time step.
\item Update the electric field in $y$ direction by half a time step.
\item Update the particle positions in $z$ direction by half a time step.
\item Update the electric field in $z$ direction by a full time step.
\item Update the particle positions in $z$ direction by half a time step.
\item Update the electric field in $y$ direction by half a time step.
\item Update the particle positions in $y$ direction by half a time step.
\item Update the electric field in $x$ direction by half a time step.
\item Update the particle positions in $x$ direction by half a time step.
\item Update the $\vec A$ field by half a time step.
\item Update the particle momenta by half a time step.
\end{enumerate}

We have added a Langevin thermostat to the particles:
\begin{equation}
\frac{d}{dt} \vec p_i = - \frac{\partial U}{\partial \vec r_i}
                        + e_i \vec E (\vec r_i) 
                        - \frac{\zeta}{m_i} \vec p_i
                        + \vec f_i ,
\end{equation}
where $\zeta$ is the particle friction constant, and $\vec f_i$ is a
random force satisfying the standard fluctuation--dissipation theorem:
\begin{equation}
\left< f_i^\alpha (t) f_j^\beta (t^\prime) \right> =
2 \zeta k_B T \delta_{ij} \delta_{\alpha \beta}
\delta (t - t^\prime),
\end{equation}
where $\alpha$ and $\beta$ denote Cartesian indices. This puts the
system into the canonical ensemble. For large systems, one can rely on
the equivalence of ensembles, and there is no fundamental
statistical--mechanical need for such a thermostat --- it is just a
matter of technical convenience: Usually a Langevin thermostat tends
to stabilize the simulation due to its inherent feedback mechanism,
such that larger time steps are feasible. Rottler and Maggs
\cite{maggs_prl_2} also add a Langevin thermostat to the magnetic
field; we have not done this. It should be noted that such
thermostatted dynamics violates time reversibility and phase--space
volume conservation anyways.

\end{appendix}

\section*{References}

% \bibliographystyle{tryiop}
% \bibliography{memd}

\end{document}